# Does God Play Dice with Universe: The Hydrogen Atomic model of Bohr and de Broglie

Pavel S. Kamenov,

Faculty of Physics, University of Sofia, Sofia-1164, Bulgaria

**In memory of Luis de Broglie**

## 1. Introduction.

In the last years some scientists think "how good is Niels Bohr's atomic model?" [1]. My paper is restricted only to the above mentioned subject and does not pretend to make any review of theory of "Hidden variable" and other phenomena. I quote only these papers which I think are useful for understanding my work.

A. Einstein wrote: "Some notions, which have proved useful for the classification of things, have gained such an authority among us that we often forget about their earthly origin and take them for unalterable facts. They are categorised as "mental principles", "given a priori". It is because of such delusions that the road of scientific progress often remains impassable for a long period of time. Therefore, it is all but futile if we exercise a little in analysing the freely used notions and identifying the circumstances which determine their validity; of how each one of them was derived from the experimental facts. Thus we will shake their excessively high authority. These concepts will be rejected if they fail to properly legitimate themselves, corrected if their adequacy to the given circumstances is too uncertain, or replaced by others if a new system, which we would for some reasons prefer, is to be set up." (Annalen der Physik, 51 (1916), p. 639-642. A free translation from German).

We hope to remind that all quantum laws were initially derived from the results of experiments with statistical ensembles of quantum systems. Subsequently these laws were applied to solitary quantum systems which are the elementary constituents of the statistical ensemble. This is easy and trivial. Easy because it is not necessary to search for other properties of the solitary object and trivial because this transition does not contradict the laws which govern an ensemble of identical objects (quantum systems, QS). For a statistical ensemble of quantum systems the introduction of probabilities and the statistical interpretation of results are inevitable, but it is not sure that a solitary quantum system must be governed by the same principles. To be more specific, I can explain the above assertions with the help of an example:

***The law of radioactive decay,*** $N=N_0\exp(-t/\tau)$, was at first observed experimentally and after that derived from statistical considerations. N is the number of nuclei which have not decayed for the time t; $N_0$ is the number of nuclei at the initial moment of time (t=0) and $\tau$ is the mean life time of all nuclei. This law can not be affected by any external interactions and is an universal one - it concerns all decays of any excited states of nuclei, atoms, molecules and so on. It is very easy to transfer this law from a statistical ensemble to one solitary object by introducing the probability (W) that this object does not decay for a time (t): $W=N/N_0= \exp(-t/\tau)$. But this probability is a trivial application of a law which concerns only a statistical ensemble of quantum objects. This probability is not a proof that a solitary object does not have another cause for decay. If such a cause exists, it will be a "hidden variable". However someone insist that hidden variables in quantum physics do not exist..

Many of the authors think that the theory of Bohr about the hydrogen atom is not adequate to reality. Bohr himself thinks (and writes) that, though his model of hydrogen describes some properties of this atom very well, it can not be accepted as realistic... Now it is considered to be ***almost a shame*** to think about this model as corresponding to reality. Some textbooks for students discuss this model of Bohr, but only as a curious example of an inadequate model which can lead to some true results [2]...

In this paper *I hope to show that this understanding has no foundations.* It stems from some mystic assumption about quantum physics [3]...For example, the complementarity principle postulated by Niels Bohr [4] in 1927, assumes that the wave-particle "duality" is a property of a single quantum system and therefore its two complementary aspects cannot be observed simultaneously; in some experiments (interference, diffraction) wave properties are manifested, some other experiments provide evidence for the corpuscular properties of quantum objects (impulse, energy). It is impossible to observe the two phenomena simultaneously and, moreover, they *cannot exist simultaneously* ...

In the paper [5] it was shown that in the case of waves on the surface of a liquid the floating classical particles which pass through only one of the two opened slits are guided by the interfering surface waves in the same directions (angle θ) as predicted by quantum laws ($|\Psi(\theta)|^2$=max; and the directions $|\Psi(\theta)|^2$=0 are not allowed.) This is an indirect confirmation of de Broglie's ideas that *wave and particle exist simultaneously and that this coexistence is real* [6]. Most of the scientists think that the field of de Broglie is not real and they accept the statistical interpretation of Born [7]. *One of the most often stressed* *disadvantages of the model of Bohr is the impossibility to determine (calculate) the probabilities of transition (intensities) of the emitted hydrogen lines...*

**2. Return to the real unitary field-particle of de Broglie and to Bohr's model of hydrogen atom.**

I ask myself why the assumption that a wave-particle can not exist simultaneously is more real than de Broglie's assumption that they *always exist simultaneously* [6]?. I hope that the results of this work will show that there is nothing more realistic than the *simultaneous existence* of de Broglie's field and Bohr's atom ... ***and that (for one atom) no statistical interpretation is necessary***. De Broglie's waves in the hydrogen atoms are such that in the stationary state the mass of the electron (m), its velocity $v_n$, and average radius $r_n$ are related with the principal quantum number (n= 1,2,3....) according to:

$$mv_nr_n = n\frac{h}{2\pi} = n\hbar \qquad (1)$$

and the field-particle (electron) is in a potential well which guides the electron only in orbit *n*, in which case the electron does not emit photons. The length of de Broglie's wave $\lambda_n$ exactly satisfies the condition:

$$2\pi r_n = \frac{nh}{mv_n} = n\lambda_n \qquad (2)$$

De Broglie's unitary "wave-particle" is in a stationary ("steady") state which does never change. The "wave-particle" electron is bound together with the "wave-particle" proton by electromagnetic forces and de Broglie' wave. They interfere and remain in their potential well (position) forever, like classical particles on the surface of a liquid [5]. The field of de Broglie is so real and this reality is so strong that the electromagnetic forces can not destroy this interference and the field-particle (electron) can not emit a photon-soliton [8,9]. In order to explain the decay of a stationary state it is necessary to assume some infinitely small "external perturbation" which would disturb the exact equalities (1) and (2) and (after some time of destructive interference) permit the transition to lower states. Only the ground state ($2\pi r_1 = \lambda_1$; n=1) can not be disturbed by an "infinitely small perturbation" because the field-particle can not be destroyed (n can not be smaller than 1). In this case only if the perturbation energy is *sufficiently* great, the electron can make a transition from the ground to the upper levels (absorption only, [10]).

Excited by some energy, the electron can randomly occur at any distance ($r_{ni}$) *around* the exact radius of the stationary orbit ($2\pi r_{ni} \approx n\lambda_n$;). Despite that the difference between the trajectory of the electron ($2\pi r_{ni}$) and $n\lambda_n$ can be very small, destructive interference leads (after some time) to a transition to lower states. We can imagine that the "wave-particle" electron self-interferes as long as the minima of the wave coincide with the maxima of the precedent waves (this means that the amplitude (D) of the interfering electron-wave becomes $|D(t)|^2 = 0$). In this moment a transition occurs and energy is emitted. The greater the difference $|r_{ni} - r_n|$, the smaller the time necessary for destructive interference. If $|r_{ni} - r_n| \to 0$, the time for destructive interference would be very long [11]. When the energy of excitation is exact ($r_{ni}=r_n$), a true stationary state would be established and without external perturbation this state could not be changed. So, it is evident that the wave-particle electron can be excited so, as to occur at all possible distances r from the proton (except $r<r_0$; ground state).

### 3. Own life time of a single hydrogen atom.

In Fig.1 a schematic wave-particle in some excited state of the hydrogen atom is shown. The particle-wave electron moves from left to right (for example, n=2). In Fig.1 a) the velocity of the electron $v_n$ is such that $\lambda_n$ and $r_n$ correspond exactly to Bohr's conditions:

$$\lambda_n = \frac{h}{mv_n} \qquad (3)$$

Such a wave-particle electron returns from the left always with the same phase and reiterates its motion for an infinitely long time. If the velocity of the electron (v) is slightly different, the new $\lambda$ will also be slightly different (compared with $\lambda_n$):

$$\lambda = \frac{h}{mv} \qquad (4)$$

Such a particle-wave electron would arrive from the left (Fig. 1 b)) with a slightly different phase. With time this difference increases and we can calculate the moment when the sum of two amplitudes becomes zero (for the first time). In this moment the electron is not more in the potential well of the wave (like classical particles, [5], when $|\Psi|^2=0$) and is allowed to change its position. This moment will be the time of life of this excited atom. The sum of the two amplitudes of de Broglie' field (D) can be written (like with classical particles, [5]):

$$D = \sin\left(\frac{2\pi}{\lambda}(vt - r)\right) + \sin\left(\frac{2\pi}{\lambda}(vt)\right) \qquad (5)$$

where r is the new radius which is only slightly different from $r_n$. The relation between $\lambda$, $\omega$ and v is:

$$\lambda = \frac{2\pi v}{\omega} \qquad (6)$$

One can substitute this in (5) and obtain:

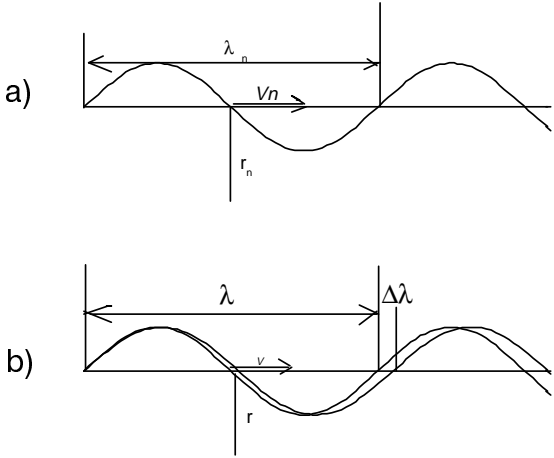

Fig. 1. A scheme of one of the first hydrogen excited states. Wave-particle electron and its interference; a) true stationary state; b) an almost stationary state.

$$D = \sin\left(\omega t - \frac{\omega r}{v}\right) + \sin(\omega t) \qquad (7)$$

Because of Bohr's model, $r/v = 1/\omega$, and (7) becomes

$$D = \sin(\omega t - 1) + \sin(\omega t) \qquad (8)$$

which is the sum of de Broglie's amplitudes (D), expressed by the time and the frequency of a not exactly stationary state. From (3) and (4) one can find the small difference $\Delta\lambda$ and $\Delta\omega$:

$$\frac{\Delta\lambda}{\lambda} = \frac{\lambda_n - \lambda}{\lambda} = \left(\frac{v}{v_n} - 1\right) = \frac{\omega}{\Delta\omega} \qquad (9)$$

Taking into account that in Bohr's model

$$\frac{v}{v_n} = \sqrt[3]{\frac{\omega}{\omega_n}} = \sqrt[3]{\frac{\omega_n + \Delta\omega}{\omega_n}} \qquad (10)$$

and from (9) we obtain ($\omega$):

$$\omega = \left(\Delta\omega\left(\sqrt[3]{\frac{\omega_n + \Delta\omega}{\omega_n}} - 1\right)\right) \qquad (11)$$

As usual, we must find the moment (t) when $|D|^2 = 0$ (the electron is not in the potential well of its wave):

$$|D|^2 = |\sin(\omega t - 1) + \sin(\omega t)|^2 = 0 \qquad (12)$$

Hence

$$\sin(\omega t - 1) = -\sin(\omega t) \qquad (13)$$

or

$$\omega t - 1 = -\omega t; \qquad t = \frac{1}{2\omega} \qquad (13a)$$

So, substituting ω from (11), we find the necessary "own lifetime" (t):

$$t = \frac{1}{2\Delta\omega\left(\sqrt[3]{\frac{\omega}{\omega_n}} - 1\right)} = \frac{1}{2\Delta\omega\left(\sqrt[3]{\frac{\omega_n + \Delta\omega}{\omega_n}} - 1\right)} \qquad (14)$$

As it is seen, when ω=$\omega_n$ (or Δω=0), the time is t→∞, as it should be for a stationary state. For Δω<<$\omega_n$, the expression for the time (14) is symmetric (for positive and negative Δω). It is more convenient (for me) to transform eqs.(14) in terms of energy:

$$t = \frac{\hbar}{2\Delta E\left(\sqrt[3]{\frac{\Delta E}{E_n} + 1} - 1\right)} \qquad (15)$$

where the energy can be measured in units eV and ℏ [eV.s]. In this case the energy ($E_n$) of the different excited states can be expressed through the Rydberg constant (R). Thus, the life time of each single excited hydrogen atom depends on the small energy difference (ΔE) and the principal quantum number (n):

$$t = \frac{\hbar}{2\Delta E\left(\sqrt[3]{1 + \frac{n^2 \Delta E}{R}} - 1\right)} \qquad (16)$$

In the case when $n^2\Delta E \ll R$, the cubic root can be expanded in a series, and taking only two first terms of the expansion $(1+n^2(\Delta E)/3R...)$ we obtain:

$$t = \frac{3\hbar R}{2(\Delta E)^2 n^2} \qquad (17)$$

Part of the results are shown in the Fig.2 (for ℏ=6.59x10$^{-16}$ eV.s and R=13.595 eV). These curves are different for different excited states (n). They could be compared with the normalized "own lifetimes" of nuclei (t/τ and ΔE/Γ) [11].

## 4. The natural width and mean life time of an ensemble of excited hydrogen atoms

Similar to the results in [11], we established that the "own life time" (t) of one single excited atom (in state (n)) depends exactly on the energy difference (ΔE) (17).

The own life time (t [s]) is determined by the exact energy of excitation ($\Delta E = |E_n - E|$), the constants of Planck ($\hbar$) and Rydberg (R), and the principle quantum number (n) of the excited state.

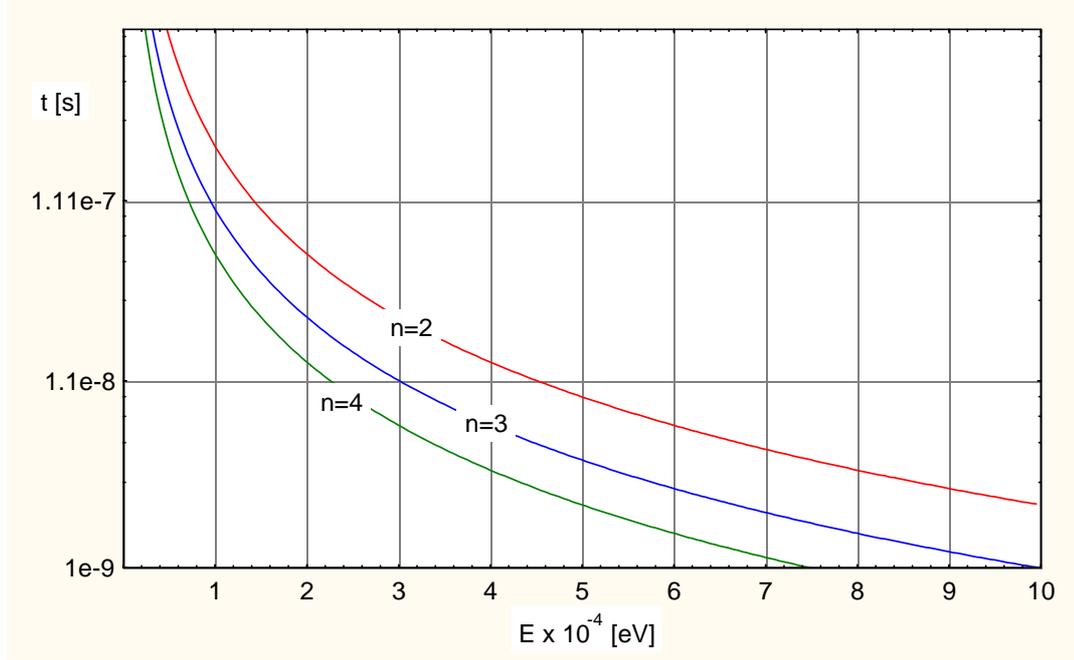

**Fig. 2**. Time (t) versus energy ($\Delta E = E_n - E$) for n=2,3 and 4. Here $E_n$=0 and these curves are symmetrical to the curves for energy differences ($-\Delta E$) (to the left of $E_n$=0). (Time (t) and energy difference (E) are in absolute units [s, eV]).

This time cannot be measured experimentally (except in the case shown in [11] for resonant Mossbauer transitions in nuclei). Experiments with hydrogen measure only the mean life time of an ensemble of excited atoms.

Further we attempt to find the statistical natural width of the levels ($\Gamma_n$) and mean life times ($\tau_n$) (for different excited states) of an ensemble of hydrogen atoms and compare these life times with reference data. Let us assume that $N_0$ [cm$^{-3}$] atoms (thin target) are irradiated by a flux of photons with uniform energy distribution $\Phi(E) = \Phi_0$ [cm$^{-2}$s$^{-1}$] = const. (in the region of some level). If the effective cross-section of excitation is $\sigma_E$, then the activity which can be obtained is:

$$\frac{dN}{dt}(t) = \Phi_0 \sigma_E N_0 \left(1 - \exp(-t/\tau_n)\right) \qquad (18)$$

As it is well known, after irradiation is terminated, activity changes with time in the following way:

$$\frac{dN}{dt}(t) = \Phi_0 \sigma_E N_0 \left(\exp(-t/\tau_n)\right) \qquad (19)$$

On the other hand, the differential cross-section ($d\sigma_E$) is:

$$d\sigma_E = \frac{\sigma_0 \Gamma_n dE}{4(\Delta E)^2 + \Gamma_n^2} \qquad (20)$$

($\sigma_0$ is the cross-section in the maximum and $\Gamma_n$ - the natural width of an ensemble of atomic levels). Then effective cross-section ($\sigma_E$) will be:

$$\sigma_E = \frac{\pi \sigma_0}{2} \qquad (21)$$

Substituting (21) in (19) we obtain the variation of activity with time after excitation:

$$\frac{dN}{dt}(t) = \Phi_0 \frac{\pi \sigma_0}{2} N_0 \left( \exp(-t/\tau_n) \right) \qquad (22)$$

Under the same conditions, but using the differential cross-section (20), we can find how activity $\frac{dN}{dt}(E)$ increases with irradiation time:

$$\frac{dN}{dt}(E) = \frac{\Phi_0 N_0 \sigma_0 \Gamma_n dE}{4(\Delta E)^2 + \Gamma_n^2} \left( 1 - \exp(-t/\tau_n) \right) \qquad (23)$$

In order to derive an expression for this activity after the end of irradiation, from (17) we obtain the variation of the own life time (t) with energy:

$$dt = \frac{3\hbar R dE}{(\Delta E)^3 n^2} \qquad (24)$$

Because of the symmetry of (17), (Fig. 2), with respect of energy, in the time interval (dt) decay the atoms in the two intervals $\Delta E = (E_n - E)$ on both sides of $E_n$:

$$dt = \frac{3\hbar R dE}{(\Delta E)^3 n^2} + \frac{3\hbar R dE}{(\Delta E)^3 n^2} = \frac{6\hbar R dE}{(\Delta E)^3 n^2} \qquad (25)$$

or

$$dE = \frac{(\Delta E)^3 n^2 dt}{6\hbar R} \qquad (26)$$

Substituting (dE) in (23) one can find the activity of hydrogen atoms (after irradiation):

$$\frac{dN}{dt}(E) = \frac{\Phi_0 N_0 \sigma_0 \Gamma_n (\Delta E)^3 n^2 dt}{\left(4(\Delta E)^2 + \Gamma_n^2\right) 6\hbar R} \qquad (27)$$

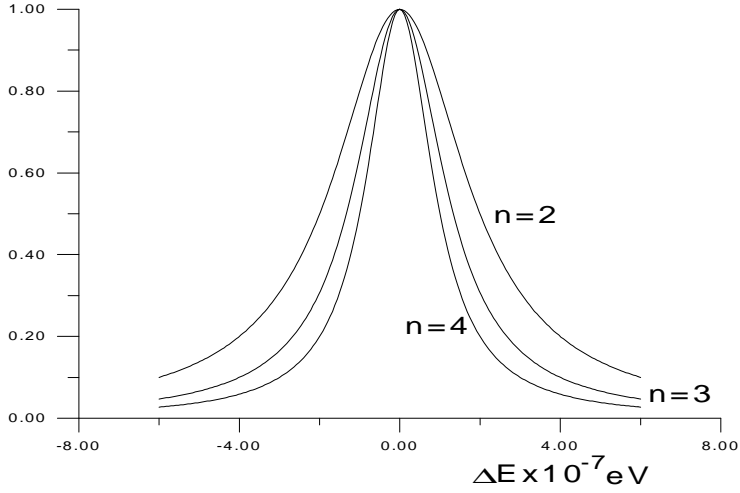

**Fig. 3**. The normalized natural lines of hydrogen atom (n=2,3 and 4). The energy ($\Delta E$) is calculated in absolute units [eV].

Thus we obtain two expressions for the activities: (27), depending on the energy of excitation ($\Delta E$), and (22), depending on time (t). Using the previous results from [11], we require that the two activities (22) and (27) be equal:

$$\frac{\Phi_0 N_0 \sigma_0 \Gamma_n (\Delta E)^3 n^2 dt}{\left(4(\Delta E)^2 + \Gamma_n^2\right) 6\hbar R} = \Phi_0 \frac{\pi \sigma_0}{2} N_0 \left(\exp(-t/\tau_n)\right) \qquad (28)$$

In the specific case (Fig. 3 and 4) when $\exp(-t/\tau_n) = 1/2$, then $\Delta E = \Gamma_n/2$, and the expression (28) becomes:

$$\frac{\Gamma_n^2 n^2 dt}{24 \hbar R} = \pi \qquad (29)$$

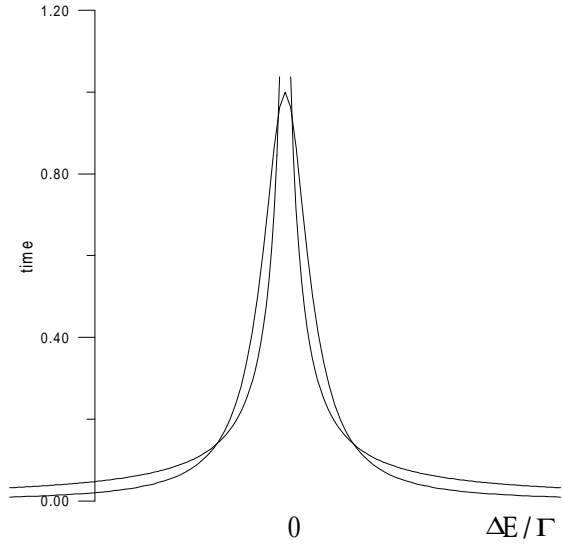

**Fig. 4**. The time (t/τ) and normalized natural lines of hydrogen atom. The energy is (ΔE/Γ). The maximum of the Lorentzian is at ΔE=0, where (t/τ) →∞. When ΔE=Γ/2 one half of the Lorentzian population of level would decay.

Hence, the natural width ($\Gamma_n$) of a statistical ensemble of atoms (for unit time interval, dt=1) can be calculated as:

$$\Gamma_n = \frac{1}{n}\sqrt{24\pi\hbar R} \qquad (30)$$

The natural line width (normalized in the maxima) are shown in Fig. 3. From the natural width ($\Gamma_n$) of level (n) it is easy to derive the mean life time of all excited atoms (on level n):

$$\tau_n = \frac{\hbar}{\Gamma_n} = n\sqrt{\frac{\hbar}{24\pi R}} \qquad (31)$$

Thus, for calculation of the full mean life time of an excited hydrogen level (n), only Rydberg's constant (R) and Planck's constant ($\hbar$) are needed. The corresponding decay constant (the spontaneous coefficient of Einstein) is $A_n = 1/\tau_n$.

## 5. Comparison with reference data.

In the numerous reference tables on hydrogen I found, to my great surprise, quite different values for $\tau_n$ (especially for high excited states). In Table 1 below I quote the data from [12] (1966) and [13] (1986) and compare them with my calculations (formula 31, 1997).

So, the result from the present calculations is in excellent agreement with reference data [13] (for n=2). It is necessary to stress that my calculations fit better to the values in [13]. The differences between the values in [12] and [13] are greater than the differences between my calculations and the data in [13]. So, the *Bohr's model (complemented with de*

*Broglie' ideas) continue to describe hydrogen properties (mean life time, natural width of the levels) as exactly as Bohr's hydrogen model describes the frequency of radiation.*

### 6. Differences between the data.

As it is known, the experimental accuracy for frequency measurement is very great in comparison with accuracy of time measurements. Here I will attempt to explain the great differences between reference data (for n>3). Experimental results are very good only for the first excited states... I think that the differences between reference data (for n>3) are due both to experimental difficulties and (mainly) to the wrong application of the relation between Einstein's coefficients, which is explained in [10,14].

| n | Data sources | | |
|---|---|---|---|
|   | [12] (1966) | [13] (1986) | (1997) |
|   | $\tau_n[s]$ | $\tau_n[s]$ | $\tau_n[s]$ |
| 2 | $2.12 \times 10^{-9}$ | $1.60 \times 10^{-9}$ | $1.603 \times 10^{-9}$ |
| 3 | $1.00 \times 10^{-8}$ | $3.94 \times 10^{-9}$ | $2.405 \times 10^{-9}$ |
| 4 | $3.3 \times 10^{-8}$ | $8.0 \times 10^{-9}$ | $3.2 \times 10^{-9}$ |

**Table 1.** The values of $\tau_n = 1/A_n$ from this work (1997) are closer to the values of data source [13] (1986). The difference between the data from [12] (1966) and [13] (for n>2) are impermissible.

In [12] the transition probability for spontaneous emission from upper state k to lower state i, $A_{ki}$, is related to the total intensity $I_{ki}$ of a line of frequency $\nu_{ik}$ by

$$I_{ki} = \frac{1}{4\pi} A_{ki} h \nu_{ik} N_k \quad \text{(expression (1) on page ii of [12])} \quad (32)$$

where h is Planck's constant and $N_k$ the population of state k. It was shown in [10,14] that this relation holds for transitions from any excited state k to the ground state i only. If (i) is also an excited state, then relation (32) must be:

$$I_{ki} = \frac{1}{4\pi}(A_{ki} + \frac{g_i}{g_k} A_{ix}) h \nu_{ik} N_k \quad (33)$$

where $A_{ix}$ is the full decay constant of level (i) and $g_i$, $g_k$ are the corresponding statistical weights. Only when $A_{ix}=0$ (ground state), (33) coincides with (32). The same applies for the transition probability of absorption $B_{ik}$ and the transition probability of induced emission $B_{ki}$ in [12]:

$$B_{ik} = 6.01 \lambda^3 \frac{g_k}{g_i} A_{ki} \quad \text{(expr. (6), p. vi of [12]} \quad (34)$$

$$B_{ki} = 6.01 \lambda^3 A_{ki} \qquad \text{(expr. (7), p. vi of [12])} \qquad (35)$$

($\lambda$ is the wavelength in Angstrom units). When (i) is an excited state, these relations are also wrong. According to [10,14], these relations (in the same units as in [12]) will be:

$$B_{ik} = 6.01 \lambda^3 \left( \frac{g_k}{g_i} A_{ki} + A_{ix} \right) \qquad (36)$$

$$B_{ki} = 6.01 \lambda^3 \left( A_{ki} + \frac{g_i}{g_k} A_{ix} \right) \qquad (37)$$

*The obtained in this paper mean lifetimes of excited levels of the simplest atom - hydrogen - are in surprising agreement with the known data for n=2. At the same time, the differences between the reference values for n>2, shows that all reference data for transition probabilities in hydrogen must be critically examined and adjusted accurately according these results.*

**7. Mossbauer experiments in the energy and time domain**

It is clear that the different regions of the level's Lorentzian population must decay at exact (and different) moments of time (t), and the emitted photons must have the energies of these regions, i.e. $E_1$, $E_2$, $E_3$ (Fig. 5, a). It is very difficult to verify this fact experimentally (for hydrogen) because of two insurmountable difficulties: a) the absence of detectors with such high time and energy resolution and b) the natural line of the hydrogen atom cannot be observed experimentally (the actually emitted line is many times wider than the natural level width, $\Gamma_n$). Natural lines of emission can be observed experimentally only with nuclei which have Mossbauer transitions (between the ground and the first excited state). As it is known, Mossbauer experiments in transmission geometry are normally performed with a very thin Mossbauer source (in order to obtain a natural Lorentzian line). The emitted photons pass through a resonant absorber (with arbitrary thickness) and the transmitted quanta are registered by a detector. A specific example of an experiment with the natural width of the emitted Mossbauer line is shown in Fig.5,a) (curve 1). The relative velocity between source and absorber (v, in units $\Gamma$) changes only the resonant (nuclear) absorption and cannot change the atomic absorption. Thus, curve 1 corresponds to the case when resonant absorption does not exist (v>>$\Gamma$), and curve 2 corresponds to the case when the maxima of emission coincide with the maxima of absorption (v=0). As it must be (for a very thick absorber), the recoilless gamma-quanta in the central region ($E_3$) are almost completely absorbed (point I=0.1). At the wings of this line absorption can be neglected (curve 1 coincides with curve 2 ($E_1$)). The areas S1 (below curve 1) and S2 (below curve 2) of regions ($E_1$, $E_2$, $E_3$) are proportional to the numbers of the photons emitted in these regions. Thus, in the two cases (v>>$\Gamma$ and v=0) the ratio of the areas (in regions $E_1$) is: $S2_{E1}/S1_{E1}=1$, whereas in regions ($E_2$) this ratio is $S2_{E2}/S1_{E2}<1$. In the central regions ($E_3$) only those gamma quanta which are emitted with recoil (1-f) pass through the very thick absorber (the region below I=0.1). Recoilless quanta (f) are completely absorbed and this ratio is: $S2_{E3}/S1_{E3}=(1-f)/1$.

With the same experimental setup, but using time-coincidence techniques, Time Domain Mossbauer Experiments (TDME) [15] can be performed (Fig. 5,b). The start signal is triggered by the preceding transition, which is the moment of formation of the first excited state and the stop signal is the moment of detection of recoilless or recoil quanta which have passed through the absorber. Thus, the time between the two signals could be interpreted as the "own lifetime" of the first excited Mossbauer level. This assertion is valid only in some specific cases where the time dispersion in the absorber can be neglected.

Time domain Mossbauer experiments qualitatively confirm the concept of "own lifetimes" [11] of nuclei. At different times of coincidence (t), the widths of regions ($E_1$, $E_2$, $E_3$) will be different, provided that the time resolution of the detector ($\Delta t$) is constant [11].

At the moment $t=0\pm\Delta t$ the ratio of the numbers of coincidences for the two cases is equal to 1, which corresponds to the ratio ($S2_{E1}/S1_{E1}$). At time $t>3\tau\pm\Delta t$ the ratio of the numbers of coincidences is: (1-f), corresponding to ($S2_{E3}/S1_{E3}$). These specific time domain experiments were repeated many times after Lynch [15]. In Fig. 5,b) the solid line corresponds to the theory in [15], and the dotted line corresponds to my calculations [11].

All experimental results (crosses) in the time domain agree with the theory in [15] because the time dispersion in the absorber is taken automatically into account, whereas for

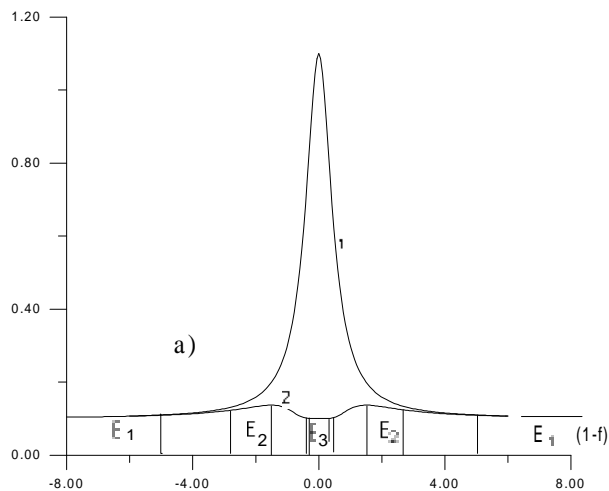

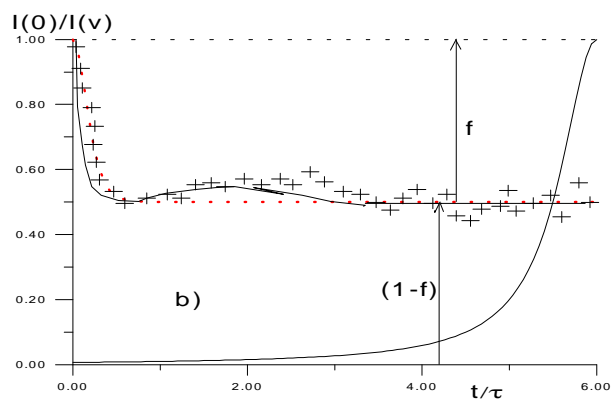

**Fig. 5**. The energy domain is at upper part a). Mossbauer lines transmitted through the absorber: no resonance (vibration; $v\gg\Gamma$) - curve 1 and resonance (v=0) - curve 2. Effective absorber thickness $D_m$=10; recoilless part f=0.5. Source and absorber - $CaSnO_3$: (source 119m-Sn, absorber 119-Sn).(see G. Hoy [18]). The recoilless part (f) is plotted above the 0.1 intensity level (I=0.1). The regions $E_1$ decay at t=0 (within the resolution time); the regions $E_3 \approx 0.024\Gamma$ decay at $t \approx 3.5\tau \pm \Delta t$.

b) A time domain transmission Mossbauer experiment. The intensity is the ratio (I(0)/I(v)); $\tau = 26.10^{-9}$ s (the resolution time is $\Delta t \approx 0.05\tau$). The crosses represent the experimental points. All other parameters are the same as in a). The solid line is calculated after the theory of Lynch [15]; the dashed line represents our calculations (not taking into account the time dispersion in the absorber). After $3.5\tau$ the experiments (like all other known experiments) and the two calculations coincide. (For f=1, see also Hoy [18]). The line with maximum at $t/\tau$=6 is only a half from a Lorentzian. Fare from resonance decay is at t=0, and about resonance $t/\tau \geq 3.5$.

the dotted line this is not true. All such *specific*

experiments show that first decay the nuclei in regions $E_1$ (t=0±Δt)), and the nuclei of regions $E_3$ decay long after those in regions $E_1$. These experiments show that for Mossbauer transitions the own lifetime of the nucleus depends on the energy difference $\Delta E = E_r - E$ (like for the hydrogen atom). The energy spectrum (Fig. 5 a) was confirmed in the excellent experimental work of Mandjukov et al. (with the help of resonance detector, [17]).

*I think that these experimental results (comparison between time and energy domains) cannot be explained with other phenomena except the "own lifetime"*. This conclusion is inevitable, unless one resorts to some mystic assumptions (for example that recoilless (f) and recoil (1-f) nuclei decay at different moments after the excitation).

**8. Some inevitable conclusions.**

The main result of this work (and works [11]) is that the first excited states of a quantum system decay after some exactly predictable time (t) according to (17). Decay is not an accidental event as it is believed by the majority of scientists (except Einstein who wrote that a weakness of the theory of radiation is that the time of occurrence of an elementary process is left to "chance" [16]). *The mean life time ($\tau_n$) is a characteristic only of a statistical ensemble of excited atoms.*

I understand that now it is very difficult to accept the concept of solitary quantum systems (SQS). In present days the majority of scientists believe that quantum physics is a fully completed science. We know all about quantum physics, and what we know are all properties of nature. So, we could not hope to achieve a better understanding of nature because there are no hidden variables. Nature is governed by probabilities, uncertainties, duality, and so on...

A referee of "Bulgarian Journal of Physics" wrote the shortest *negative* reference on my paper about the solitary hydrogen atom (this paper, 1997): "This paper cannot be published in the Journal because the author writes that an excited state decays at an exactly predictable moment of time". This shortest reference was sufficient for the editorial board to reject the paper... The Editor (my friend, M. M.) tells me the same expression ...

And yet another example: The Editor of Physics Letters A wrote (more than the first): "...Despite my considerable efforts I have been unable to elicit any referee reports on your article and at this stage I do not think I am going to receive any reviews. In the circumstances I believe it would be in your interests if you submitted your article to another journal. I appreciate this is not a very satisfactory outcome but wish you luck in publishing your work elsewhere". (The paper was written in 1997!).

This situation resembles a journey through the jungle to a city to which there are numerous excellent roads. If someone of the travellers decides to cross straight through the jungle, his fellows will try to stop him with the words: "You do not have to do this. All roads lead to our city. It is dangerous and stupid to force your way through the forest and the shrubs when you may take any of the beaten tracks…" However, they will never know if amidst the jungle there exist other smaller towns or villages which may some day grow into large beautiful cities…

This article is a step aside from the beaten tracks. Such steps may prove to be illusory or wrong, and none of them are the first ones in the corresponding directions. The first steps were made by other people (mentioned in the paper). While some of the tracks may eventually lead nowhere and never become useful roads, all roads are made by numerous travellers, each of them stepping slightly aside from his predecessor. Only when the first one to pass is followed by many others, it is possible to open a new road. And, to be sure, each road links places which are useful to people. As one Bulgarian poet (Penyo Penev) wrote: "All roads lead to the men".

If the first traveller does not find something useful, if he fails to explain its advantages, or if there is *no one to understand him*, the first steps remain lonely for a long time. But, of course, where useful steps have been made, sooner or later there will be a road! Such is human nature, and one can feel pity for all those who suffer from the deeply rooted illusion that only beaten tracks can lead us to something important.

I hope that some of the results presented in this article may shake the exaggerated authority of the existing concepts not only in quantum physics, but also in classical physics. I am convinced that a great many of the delusions in quantum physics originate from our wrong notions about nature, established as irrefutable truths in classical physics already at the time of Newton and Galilee. The very separation of science into "classical" and "quantum" speaks in support of this. There is only one physics, just like there is only one Nature. We simply study the nature of things (the jungle) still deeper and deeper… and in more details. A Spanish proverb tells: "Traveller, roads do not exist by themselves. They are made by travelling."

This work is a part of my unpublished book entitled: **"Physics of Solitary Quantum Systems - From Dice to Chronometers"**. It is shown in the book that God does not play dice with solitary quantum systems like hydrogen, nucleus, photon. God does not play dice with Universe ... I am happy that now (in the last years before 2000) there exists the possibility for publishing such papers without a referee with deeply rooted illusions ...

*Acknowledgment.* This work was supported in part by the Bulgarian National Foundation for Scientific Research (No 534/1995).